\documentclass[preprint,showpacs,preprintnumbers,amsmath,amssymb]{revtex4}
\usepackage{graphicx}
\usepackage{dcolumn}
\usepackage{bm}

\begin{document}

\title{Measuring entropy generated by spin-transfer} 
\author{J.-E. Wegrowe} \email{jean-eric.wegrowe@polytechnique.edu} 
\author{Q. Anh Nguyen, T.L. Wade} 
\affiliation{Laboratoire des
Solides Irradi\'es, Ecole Polytechnique, 91128 Palaiseau Cedex, France.}

\date{\today} 

\begin{abstract}
An experimental protocol is presented that allows the entropy
generated by spin-transfer to be measured.  The effect of a strong
spin-polarized current injected on a ferromagnetic nanostructure is
investigated with focusing on the quasi-static equilibrium states of a
ferromagnetic single domain.  The samples are single contacted Ni
nanowires obtained by electrodeposition in a nanoporous template.  The
thermal susceptibility of the magnetoresistance is measured as a
function of the magnetic field for different values of the current
injected through the wire.  This quantity is related to the thermal
magnetic susceptibility of the ferromagnetic wire through the
anisotropic magnetoresistance.  The ferromagnetic entropy generated by
the current injection is deduced thanks to a thermodynamic Maxwell
relation.  This study shows that the effect of the spin-transfer in
our samples results in the generation of incoherent excitations
instead of rotation of the magnetization.
\end{abstract}

\pacs{75.40.Gb,72.25.Hg,75.47.De \hfill}

\maketitle

Spin-transfer is the generic name for the effect of magnetization
reversal or magnetic excitations produced by the injection of a
spin-polarized current in a ferromagnetic nanostructure.  Since the discovery
of spin-transfer effects in various kinds of systems \cite{liste0},
some fundamental questions emerged about the stochastic vs. 
deterministic nature of the effect.  Already in the
pioneering theoretical works about spin-transfer, the two
approaches were suggested.  Luc Berger \cite{Berger} proposed first a
non-deterministic approach with calculating an electronic s-d mean
relaxation time that defines the transfer of spins from the electric current to
the lattice (a consequence of which is to generate spin-waves).  On the
other hand, Slonczewski proposed a determinist term to be added to the
Landau-Lifshitz-Gilbert equation of the dynamics of the ferromagnet 
\cite{Slon} (this term is the expression of a torque exerted on the 
ferromagnet).  The deterministic term is of "dissipative " nature, 
however, in the case of homogeneous magnetization, the
corresponding effective fields still derive from two deterministic
potentials \cite{JPhys,Serpico}. The two approaches are actually 
difficult to discriminate because the deterministic terms may also 
justify the existence of incoherent excitations, which eventually lead to the 
magnetization reversal.

In the years that followed this discovery, many different aspects of
spin-transfer have been investigated experimentally, especially in
terms of critical currents necessary to switch the magnetization (at a
given excitation time scale).  However, the important point that
motivated this work is that the measure of the critical current is a
measure of the magnetic relaxation, so that it involves the stochastic
process describe by the N\'eel-Brown activation \cite{JPhys}.  The
consequence is that the process measured by the critical current is
stochastic, whatever the nature of the underlying mechanism, and after ten
years of spin-transfer measurements it is still difficult to
understand the very nature of the underlying magnetic process. 
Should
spin-transfer be described by a deterministic expression in the
ferromagnetic dynamic equation, or should spin-transfer be described
by diffusion or fluctuations in the corresponding stochastic equation? 
Or even more basically: does spin-transfer produce coherent rotation
or precession of the magnetization (without the help of an external AC
excitation), or incoherent excitations?  The ambition of this work is
to answer these questions.

It is worth pointing out that despite an intense research activity on
this topic, the simplest experiments that would avoid the activation
process, namely measuring the rotation of the magnetization due to
current injection on the {\it quasi-static equilibrium states}, i.e.
measuring the modification {\it of the reversible part of the
hysteresis loop}, has not been reported as such.  This situation is
rather surprising since it would be the most direct way to obtain the
deterministic terms predicted by Slonczewski and to exhibit the
corresponding potentials.

One problem of the quasi-static measurements is due to the
difficulty of measuring precisely the magnetization states of individual single
ferromagnetic domains (contacted to a current source) as a function of 
the electric current densities, for very high densities injected.  
Adding to the difficulty of measuring the magnetization states at the 
nanoscale, many
collateral effects of thermodynamic nature take place.  Joule heating is of
course superimposed on the effect of the spin-polarized current
injection, but also induce fields and some important thermoelectric 
effects \cite{Fukushima}.

The aim of this paper is to present a new experimental protocol that
allows the effect of spin-injection to be investigated on the
quasi-static states of the magnetization, to present the results
obtained, and to discuss the results in terms of ferromagnetic
fluctuations generated by spin-transfer.  This goal is achieved thanks
to remarkable thermodynamic properties of ferromagnets.  In analogy
with magnetocaloric effects, we show that thermal susceptibility
measurements of the anisotropic magnetoresistance of a single
ferromagnetic domain give access to the entropy production, and that
this method furnishes a sensitive probe of the ferromagnetic
fluctuations generated by spin-injection (i.e. spin-transfer).  It is
interesting to recall that such an approach was used for the first
experimental evidence of the spontaneous magnetization of a
ferromagnet as well as the first check of the mean field theory
\cite{Weiss}.

The structure used for this investigation is a single contacted
Ni nanowire, obtained by electrodeposition into
nanoporous polycarbonate membranes.  This sample is chosen for convenience as
typical test system thanks to previous studies related to the
magnetization states \cite{WW,PRLMoi,PRBYvan,JMMM200}, to
spin-transfer measured for the irreversible jumps of the magnetization
\cite{EPL,Guittienne,PRBKelly,Dubey}, and spin-dependent thermoelectric
power \cite{Gravier,Santi,MTEPW,Shapira}.  The nanowire can be described 
as a
single ferromagnetic layer with two ferromagnetic/normal
interfaces \cite{JPhys}.  The magnetization is uniform for all equilibrium states:
$\vec M = M_{s}.  \vec u_{r}$, where $M_{s}$ is the magnetization at
saturation, and $\vec u_{r}$ is the radial unit vector.  In contrast
to the spin-valve systems, the ferromagnetic layer
is large with respect to the spin-diffusion length (and to other typical
relaxation lengths), i.e. large with respect to the interfaces.  This
system is simple enough to discriminate easily between bulk and
interface effects, which is the very first step in the understanding
of the measured effects (see references \cite{MTEPW,Santi}).
The protocol described here can also be applied as such on typical pillar 
spin-valves and tunneling junctions,
e.g. the memory units used for Magnetic Random Access Memory (MRAM) applications.

For nanowires composed by a uniform ferromagnetic layer, the
magnetization is measured through the anisotropic magnetoresistance
(AMR).  The giant magnetoresistance (GMR) cannot be measured directly
with one magnetic layer because there is no reference states for
spin-flip scattering \cite{JPhys}.  Nevertheless, spin-accumulation
(responsible for the GMR in spin-valve structures) is present at the
interfaces between the normal metal and the Ni in both sides of the
wire \cite{MTEPW}, and spin-transfer is clearly observed
\cite{Myers,EPL,Guittienne,ChenPRL,PRBKelly,Dubey}.

The experimental protocol adopted here in order to access to the
effect of spin transfer on the reversible part of the hysteresis is
the following: the thermal susceptibility of the magnetization
$\chi_{T} = \frac{1}{M_{s}} \frac{\partial M}{\partial T}$ (where $T$ is
the temperature of the wire and $M_{s}$ its magnetization at saturation)
is measured through the variation of the AMR in phase with the
temperature. The measurements are performed as a
function of the external field, and as a function of the amplitude of
the current injected.

The field dependent thermal susceptibility allows the ferromagnetic
entropy production to be obtained, according to the Maxwell relation:
$\left ( \frac{\partial S}{\partial H} \right )_{T} = 
\left ( \frac{\partial M}{\partial T} \right )_{H}$. The validity of the Maxwell
relation is assumed because the total magnetic system is
considered (a closed
ferromagnetic system that takes into account the spin-dependent
current generator and the ferromagnetic layer is defined in the references
\cite{JPhys,PRB08}), and because only the reversible part of the
magnetization curve is taken into account (see \cite{Basso} otherwhise).
  
The paper is composed as follows: the first part is devoted to the
characterization of the ferromagnetic states, with the use of the AMR
properties of the Ni nanowires.  The second part presents the
measurement protocol and the relations that link the measured thermal
susceptibility to the fluctuations and to the usual field susceptibility
$\chi_{H} = \frac{1}{M_{s}} \frac{\partial M}{\partial T}$ deduced 
from the AMR.  The third
part shows the action of high current densities on this
susceptibility and the entropy generated by 
spin-transfer. We conclude that the effect of the spin-transfer in our 
samples results in the generation of incoherent excitations.

\subsection{AMR and the reversible part of the hysteresis}

The interest in working with Ni nanowires is that the ferromagnetic
states are well known and are easy to describe
\cite{PRLMoi,PRBYvan,PRBKelly,JMMM200}.  The wires length is $l = 6$
$\mu$m and the diameters is about $60$ nm (aspect ratio 100).  The
electrodeposited Ni is composed of very small crystallites inside the
wire, so that the magnetocrystalline anisotropy is averaged-out over
some tens of nanometers.  From the ferromagnetic point of view the
anisotropy is uniaxial, because it is reduced to the shape anisotropy
of a very long cylinder (with aspect ratio 100, the demagnetizing
factor is very close to that of an infinite cylinder).

 The
wires are contacted individually by an in-situ method inside the
electrolytic bath.  The method consists in depositing a thin gold
layer on the top of the membrane (without obstruction of the pores) in
order to measure the potential between the top and the bottom of the
membrane during the growth of Ni.  A feed-back loop allows the
deposition to be stop for a single wire contacted \cite{IEEEMoi}.  The
fabrication process is described elsewhere \cite{Travis}.  The
wire is contacted to a current source and a voltmeter.

\begin{figure}
\begin{center}
	\includegraphics[height=8cm]{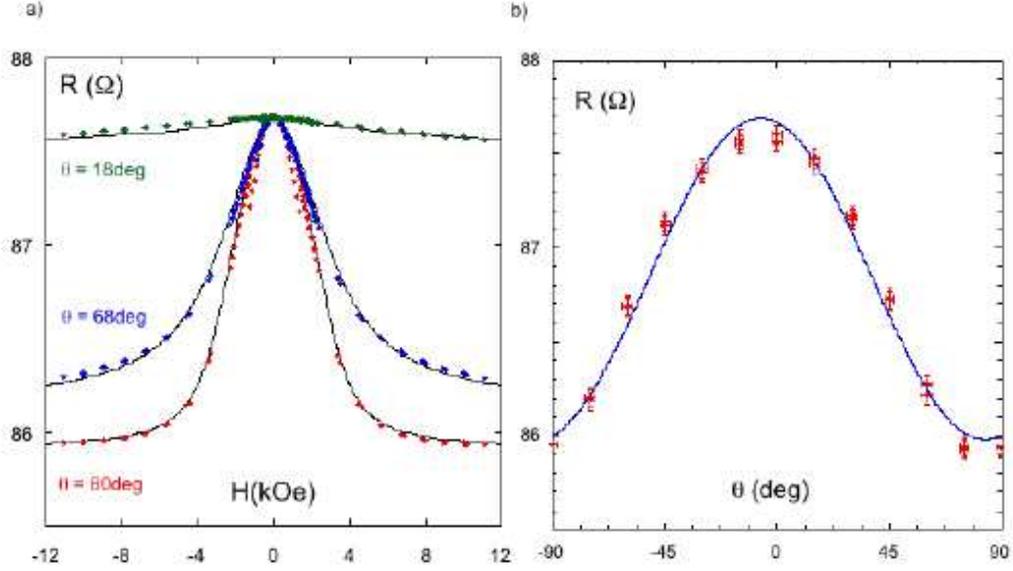}
	\caption{(a) AMR hysteresis loops as a function of the external
	field for different angles with respect to the wire axis (dots).  The
	lines represent the reversible part of the hysteresis calculated
	for a uniform and uniaxial Ni nanowire (using the potential Eq. 
	(\ref{potential}) and the magnetoresistance Eq.  (\ref{AMR})) for the 
angles $18^{o}$, $69^{o}$ and $80^{o}$.  (b)
	AMR as a function of the angle of the magnetization at saturation
	field.  The continuous line is the $cos^{2}(\theta)$ of curve of Eq. (1).}
	\label{fig:PlotAMR}
\end{center}
\end{figure}

The resistance hysteresis loops, measured as a function of the
amplitude of the magnetic field and for different field directions, are
plotted in Fig.  1 (a).  Fig.  1 (b) shows the AMR measured at
saturation field ($H = 1.2 T$) and plotted as a function of the field
direction (defined by the angle $\theta$).  At saturation field, the
direction of the field $\vec H$ coincides with the direction of the
magnetization $\vec M$ (defined by the angle $\varphi$ in the
following) so that $ \theta = \varphi$.  Furthermore, in the case of
the nanowire, the angle $\varphi$ is also the angle between the
current density $\vec J$ (oriented along the wire axis) and the
magnetization $ \varphi = (\vec J, \vec M)$.  The Fit in Fig .  1(b)
shows that $R(\varphi) = R(\varphi = 90^{o}) + \left ( R(\varphi = 0)
- R(\varphi = 90^{o}) \right ) cos^{2}(\varphi)$.  The
magnetoresistance is hence reduced to the AMR contribution only
\cite{Potter}.

As expected for a single domain ferromagnetic layer, the profile plotted
in Fig.  1(a) is composed of four reversible parts of the hysteresis
and two symmetric irreversible jumps joining the reversible
branches.  The succession of quasi-static states that defines each
branch of the hysteresis is indeed reversible.

Thanks to the wire's geometry, the maximum of the
resistivity $\rho_{\parallel}$ corresponds to the magnetization
parallel to the wire axis ($\theta = 0$) while the minimum corresponds
to the magnetization perpendicular to the wire axis $\rho_{\perp}$. 
The uniform rotation of the magnetization, $cos(\varphi(H)) =
M(H)/M_{s}$ leads to the simple expression that relates the
magnetoresitance hysteresis loop to the magnetic hysteresis loop
\cite{PRLMoi}:

\begin{equation}
\rho(H,T) = \rho_{\perp}(T) + 
\Delta \rho(T) \, \left[\frac{M(H,T)}{M_{s}}\right]^{2}
\label{AMR}
\end{equation}

where $\Delta \rho = \rho_{\parallel} - \rho_{\perp}$, and the 
measured hysteresis $M(H)$ is that obtained by projection over the wire 
axis.

On the other hand, the quasi-static states of the magnetization, that
defines the reversible part of the hysteresis loop, are given by the
equilibrium condition $\partial V/\partial \vec M = 0$
\cite{Brown}.  In the case of the Ni nanowires, the 
magnetization states at equilibrium are uniform, and the
anisotropy, reduced to the shape anisotropy, is uniaxial. 
The Gibbs energy of the Ni nanowire writes :

\begin{equation}
V(\varphi,H) = K \, sin^{2}(\varphi) - M_{s} H cos(\theta - \varphi)
\label{potential}
\end{equation} 

where $H$ is the amplitude of the magnetic field applied at an angle $\theta$ and
$\varphi$ is the angle of the magnetization with respect to the wire
axis.  Introducing Eq. (\ref{potential})  into the equilibrium 
conditions gives the reversible part of the hysteresis: 

\begin{equation}
K sin(2 \varphi) - M_{s} H sin(\theta - \varphi ) = 0
\label{equilibre}
\end{equation} 

The curves plotted in Fig.  1(a), calculated for the three angles 
$18^{o}$, $68^{o}$, and $80^{o}$, are obtained by using Eq. 
(\ref{equilibre}) including the AMR 
formula (\ref{AMR}) (a first fit is performed at $80^{o}$ for adjusting 
the resistance and the anisotropy parameter $K$, and the other curves are
deduced without adjusting parameters; the parameter $K$ is that used in 
the curve plotted in Fid. 1(b)). The validity of this
description is confirmed by the calculated curves Fig.  1(a) and Fig. 
1(b). The anisotropy 
field is found to be $H_{a} = 3 kOe$. This is the value expected for 
the anisotropy
field  of an infinite cylinder $H_{a} =
\mu_{0} M_{s}/2 = 3,05$ $kOe$.

\subsection{Measure of the thermal susceptibility}

In order to measure the thermal susceptibility, a Joule heater is
fixed on the bottom of the membrane (see Fig.  2).  The membrane is 
hence composed of the heater on the bottom with a $250$ nm gold layer, and a
submicrometric Ni mushroom that forms the contact on the surface of
the second gold layer of about 50 nm thickness.

\begin{figure}
\begin{center}
	\includegraphics[height=6cm]{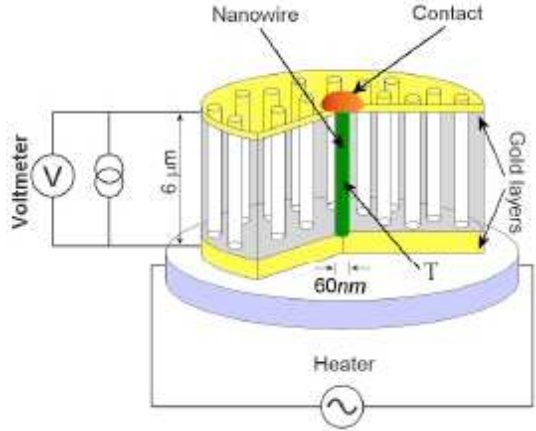}
	\caption{Schematic of the membrane with the contacted nanowire and the 
	heater.}
	\label{fig:skema}
\end{center}
\end{figure}

The analysis of the heat conductivity
regimes inside the wire \cite{Guittienne} shows that the
temperature of the membrane is not significantly modified, and the heat
propagates from the bottom layer to the top layer through the
contacted wire.

 The experimental protocol established in order to measure de thermal
 susceptibility is as follows.  A sinusoidal temperature variation is
 imposed on the wire through the heater and the response of the
 voltage (i.e. the magnetoresitance) is measured in phase.  The heater
 is controlled by a heat current oscillating at frequency f, $I_{heat}=
 I_{h0} \, cos(2 \pi f)$,
 where $I_{h0}$ is of the order of 1 mA. The heat power is dispersed
 into the whole structure, and only a small but sufficient fraction of
 the power is used to heat the wire.  The temperature inside the wire
 is measured precisely with the resistance of the wire itself 
 (typically, the 
 resistivity is of the order of $5.10^{-8}$ $\Omega.m$ and the 
 variation due to the heater is about $1$ \%). The corresponding temperature 
variation $\Delta T$ of the wire  is of the order of $3$ degrees 
around the temperature $T_{0} \approx$ 300 $K$.
 At stationary regime, the mean temperature of the wire follows the 
 heater, and oscillates in time at the frequency $2f$. 
 In order to reach a well defined thermal stationary regime, the
 frequency of the signal is fixed at $f=0.05Hz$. The duration of each 
 measurement is of the order of $120$ $sec$. Since the wire reaches its 
 thermal equilibrium much more rapidly \cite{Guittienne}, each measured 
 point is isothermic. This protocol is necessary 
 in order to extract the response of the system to the thermal 
 excitation, with the exclusion of the other contributions: drifts, 
 thermal instabilities, electromigration etc (especially while injecting very strong 
 current densities: see next section).
 The gradient of temperature between the top and bottom of the membrane
is about one degree, and is well controlledby measuring the
thermoelectric power \cite{MTEPW}.  
 
 It is important to note that the temperature of the membrane is kept
 constant (slightly above room temperature), in order to avoid
 magnetostrictive effects due to the thermal expansion of the
 polycarbonate membrane.  For
 this reason, an external (macroscopic) thermostat cannot be used in
 these experiments.
 
 During the AC temperature excitation, a current is injected in the 
 wires in order to measure the magnetoresistance. The current 
 injected at a first stage of the study is weak, typically of the order of 20 
 $\mu A$ (about $2. 10^{5}$ $A/cm^{2}$), and has no effects in terms of 
 spin-transfer. 
 The signal is analyzed numerically (using Labview program) with the
 following expression:
 
 \begin{equation}
V(t)=\textbf{\textit{a0}} + \textbf{\textit{a1}}cos(2\pi
ft+\textbf{\textit{a2}}) + \textbf{\textit{a3}}cos^{2}(2\pi
ft+\textbf{\textit{a4}})
\label{Fit}
\end{equation}
 
where:

\begin{description}
	\item[$\bullet$]	$\textbf{\textit{a0}}$ :
$\displaystyle R_{AMR} = \frac{\textbf{\textit{a0}}}I(H)$ is the 
magnetoresistance.
	\item[$\bullet$]	$\textbf{\textit{a1}}$ : correction due to the 
	induction produced by the heater.
	\item[$\bullet$]	$\textbf{\textit{a3} = $\frac{\Delta V }{\Delta T} . 
	\Delta T$ }$ : 
	amplitude of the response to the thermal excitation, or "thermal susceptibility of the 
	magnetoresistance" (this term will be justified below).
	\item[$\bullet$]	$\textbf{\textit{a2}},\textbf{\textit{a4}}$ : 
	phase corrections.
\end{description}

 The corresponding fit is plotted in the left inset of Fig.  3.  As
 will be described below, the main contribution of the signal (plotted
 as a function of the time in the inset) is due to the
 temperature dependence of the resistance $R(T)$.  The contribution of
 interest is the {\it spin-dependent contribution} of this signal that
 can be extracted from the fit (through the coefficient $a3$) and
 plotted as a function of the magnetic field $H$ (Fig.  3).  The
 profile obtained is typical for all measured samples (see also
 results and discussions in \cite{Santi,Felix,Fullerton}
 for systems exhibiting GMR).

 \begin{figure}
\begin{center}
	\includegraphics[height=8cm]{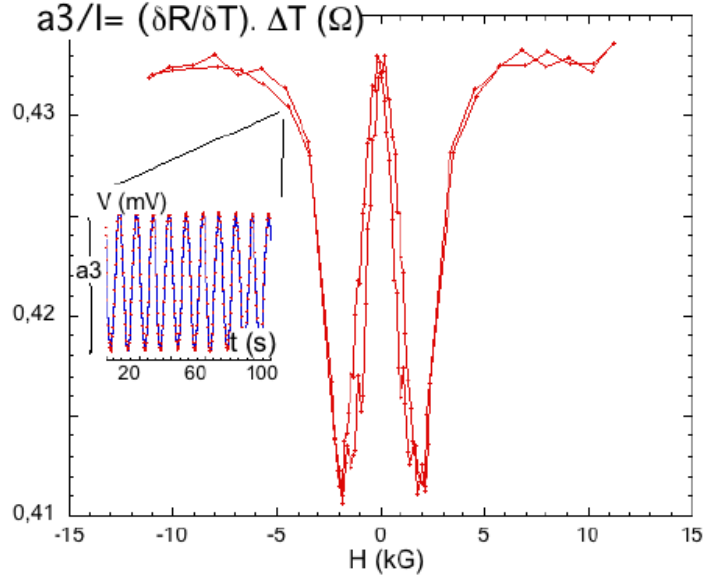}
	\caption{Thermal susceptibility of the magnetoresistance as a
	function of the magnetic field.  The angle of the applied field is
	about $80^{o}$, the current density is 0.2 $MA/cm^{2}$, and the
	amplitude of the temperature variation is $\Delta T \approx 3$ $K$
	(around $T_{0} = 300K$).  Inset: the parameter $a3/I$ is the amplitude of
	the voltage response to a sinusoidal temperature variation
	obtained by the fit (blue line).}
	\label{fig:MesurePoint}
\end{center}
\end{figure}

Figure 4 shows the profiles of the coefficient $a3/I$ $(H)$ for 
different values of the angle of the applied field. The signal
is anisotropic so that this contribution is clearly due to a
bulk effect. Interface effects (like those measured with the
thermoelectric power on equivalent samples \cite{MTEPW}) are not
observed at small current densities. The irreversible jump of the 
magnetization, observed in the AMR curve (Fig. 1) can also be seen in 
the $a3/I$ $(H)$ profile. However, we focus here on the reversible 
branches of 
the hysteresis. In the following, the angle of the applied field will 
be fixed around $80^{o}$ because the signal is large and 
spin-transfer is maximal \cite{EPL,PRBKelly}.

\begin{figure}
\begin{center}
	\includegraphics[height=8cm]{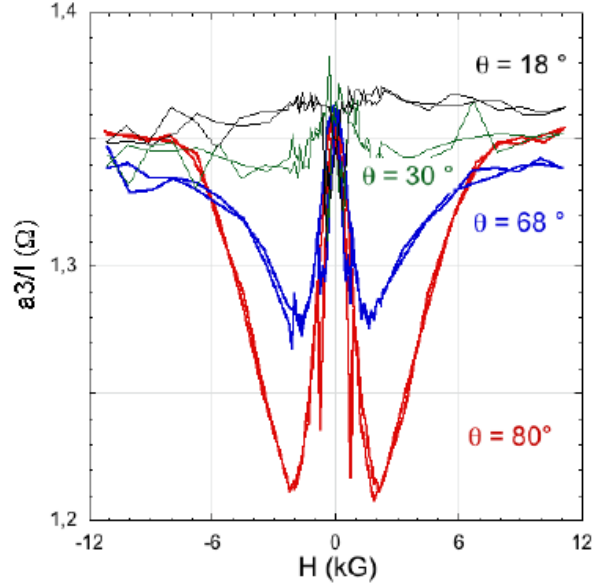}
	\caption{Magnetic susceptibility of the magnetoresistance as a 
	function of the magnetic field for different angles of the applied 
	field.}
	\label{fig:Angle}
\end{center}
\end{figure}

How can we analyze this signal? Due to the high aspect ratio, the
current density is perfectly uniform inside the wire (except at the
interface with the contacts), and from that point of view, the metallic
wire is one-dimensional.  Defining the wire axis by the coordinate $z$,
the kinetic equations writes (for our galvanostatic experiments the
current density is constant).

\begin{equation}
\label{Onsager2}
\left\{
\begin{aligned}
J_{e} &= -\frac{\sigma}{|e|} \frac{\partial
\mu}{\partial z} + \mathcal{S} \sigma\frac{\partial
T}{\partial z}\\
J_{Q} &= -\left|e\right|\mathcal{S} TJ_{e} - \kappa \frac{\partial T}{\partial z}
\end{aligned}
\right.
\end{equation}

with the electronic charge $e$ and where $\mu$ is the electrochemical
potential ($\frac{dV}{dz} = -\frac{1}{|e|} \frac{\partial
\mu}{\partial z} $ is the electric field), $J_{e}$ electric current,
and $J_{Q}$ the heat current.  The corresponding transport
coefficients are the conductivity $\sigma$ and Fourrier coefficient
$\kappa$.  The Onsager cross-coefficient are described by the Seebeck
coefficient $\mathcal{S}$ (or the Peltier coefficient $T\mathcal{S}$).  Using the
Wiedemann-Franz law $\kappa = \mathcal L_{0} \sigma T$, we introduce
the Lorentz constant $\mathcal L_{O}$.  Eq.  (\ref{Onsager2})
rewrites:

\begin{equation}
\label{gradPot}
\frac{\partial \mu}{\partial z} = -|e|\rho(H,T)\left(1 + 
\frac{|e| \mathcal{S}^{2}(T)}{\mathcal{L}_{0}}\right)J_{e}
-\frac{|e|S(T)}{\mathcal{L}_{0}T}\rho(H,T)J_{Q}
\end{equation}

where the conductivity $\rho = 1/\sigma$, function of the 
magnetization states $M$ and the temperature $T$, has been 
introduced.

The ratio $S^{2}/\mathcal L_{0} \approx 0.01$ is small for Ni
and can be neglected in the first term on the right hand side of Eq. 
(\ref{gradPot}).  On the other hand, the second term in the right hand
side corresponds to the voltage when no current passes through the
wire and gives the thermoelectric power (TEP).  This term has been
studied in details in a previous work \cite{MTEPW}.  It is here given
by the voltage measured at zero current, i.e. by the term $a3(I=0)$ according to Eq. 
(\ref{Fit}) (in the following, the coefficient $a3$ will be corrected by
the thermoelectric effect $a3(I) \leftarrow a3(I) - a3(I=0)$.  With
this correction Eq.  (\ref{gradPot}) simply reduces the Ohm's law:
$\frac{d V}{dz} = \rho(H,T) J_{e}$.  Integrating over the wire of
length $l$, we have

\begin{equation}
 V(H,T) =  \rho(H,T)\frac{l}{A} I
\end{equation}
where $I$ is the current flowing in the section $A$ of the wire. 
The validity of this expression can be check by measuring
the coefficient $a3(I)$ as a function of the current (see Fig.  5). 
The TEP, i.e. the coefficient $a3(I=0)$, is shown in the inset with
the magnification around $I=0$.  At very large $I$ (which is the
region of interest for the study of spin-transfer) the deviation
observed in Fig. 5 is mainly due to the Joule effect that modifies the
temperature of the sample.  The discussion related to spin-transfer
effect is postpone for the next section.

\begin{figure}
\begin{center}
	\includegraphics[height=8cm]{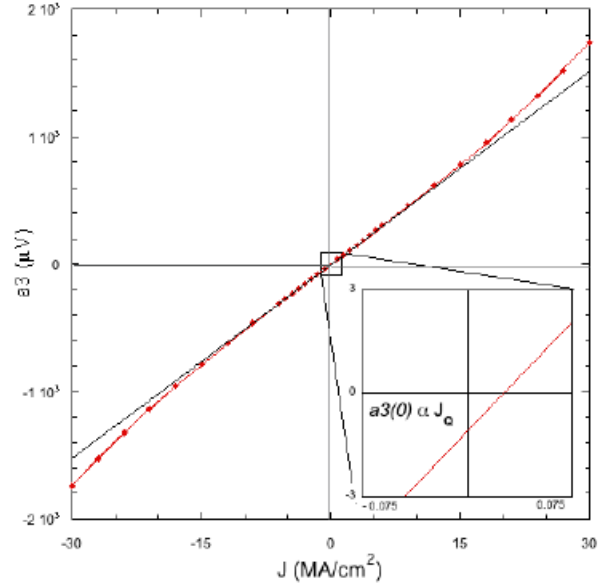}
	\caption{Thermal susceptibility of the magnetoresistance as a 
	function of the current injected into the wire at $ H=0$. The magnification 
	around $I=0$ shows the contribution of the TEP}
	\label{fig:a3(I)}
\end{center}
\end{figure}

The coefficient $a3$ measurement is defined by the thermal variation of 
the voltage. In terms of resistance we have:

\begin{equation}
\frac{a3}{I} = \frac{\partial R_{AMR}}{\partial T} \Delta T =
\left ( \frac{l \Delta T}{A} \right ) \, \frac{\partial \rho}{\partial T}
\label{a3surI}
\end{equation}

Introducing the expression of the AMR Eq(\ref{AMR})
\begin{equation}
\frac{\partial \rho(H,T)}{\partial T} = \frac{\partial
\rho_{\perp}(T)}{\partial T} + \frac{\partial \Delta \rho (T)}{\partial
T}\left[\frac{M(H,T)}{M_{s}}\right]^{2} +
2 \Delta \rho(T) \, \left[\frac{M(H,T)}{M_{s}}\right] 
\underbrace{\frac{1}{M_{s}} \frac{\partial
M}{\partial T}}_{\chi_{T}}
\label{dRdT0}
\end{equation}

where the last term defines the thermal susceptibility of the 
magnetization $\chi_{T} = \frac{1}{M_{s}} \frac{\partial
M}{\partial T} $.
In the temperature range of the experiment [290K, 310K], the temperature 
variation 
of the AMR is very small, and the second term in the right hand 
side is neglected. 
Inserting Eq. (\ref{AMR})  we obtain the thermal susceptibility as a 
function of the measured parameters :

\begin{equation}
\chi_{T}(H)  =  \frac{\partial \tilde{\rho} (H,T)}{\partial T} 
\,  \frac{1}{2 \Delta \rho (T)} \sqrt{\frac{\Delta \rho}{\tilde 
\rho(H,T)}}
\label{dRdT1}
\end{equation}

Where $ \tilde{\rho}(H,T) = \rho (H,T) - \rho_{\perp}(T) $, and 
$\frac{\partial \rho}{\partial T}$ is field independent and constant
within our temperature range. In conclusion, since both the temperature
variation of the resistance and the magnetoresistance hysteresis
loops $\rho(H,T)$ are known, the protocol described here allows the
ferromagnetic thermal susceptibility of a single nanowire to be
measured.

It should be noticed that the $a3$ profiles (Fig.  3, Fig.  4) are
rather similar to that of the derivative $\frac{\partial
\rho}{\partial H}$ of the AMR profile, i.e. the field susceptibility
$\chi_{H} = \frac{1}{M_{s}} \frac{\partial M}{\partial H}$.  This is
due to a general property: both 
susceptibilities are
proportional to the static fluctuations.  In our case, $M = M_{s} < 
cos(\varphi) >$ is
the equilibrium mean value of the magnetization (projected over the
wire axis). We have $M = 2 \pi M_{s} \, \int_{0}^{\pi} cos(\varphi)
e^{-V(H,\varphi)/kT} sin\varphi d\varphi \, /Z$, where $Z(H,T)$ is the partition
function and $V(\vec{H},\varphi)$ is composed by an anisotropy term
$K sin^{2}(\varphi)$ and the Zeeman term $- \vec{M}.\vec{H}$ (Eq. 
(\ref{potential})).  

The thermal susceptibility writes:

\begin{equation}
	\begin{aligned}
	\chi_{T} &= - \frac{H M_{s}}{kT^{2}} \, \left ( \langle cos (\varphi) 
	cos(\theta - \varphi) \rangle - \langle cos (\varphi) \rangle \langle 
	cos (\theta - \varphi) \rangle 
	\right) \\
	&+ \frac{K}{kT^{2}} 
	\left ( \langle cos(\varphi) sin^{2}(\varphi) \rangle - \langle sin^{2} \varphi \rangle
	\langle cos(\varphi) \rangle \right )
	\end{aligned}
\label{SusceptT}
\end{equation}

On the other hand, the field susceptibility writes:

\begin{equation}
	\chi_{H} = \frac{M_{s}}{kT} \, \left ( \langle cos (\varphi) 
	cos(\theta - \varphi) \rangle - \langle cos (\varphi) \rangle \langle 
	cos (\theta - \varphi) \rangle 
	\right) 
\label{SusceptH}
\end{equation}

The following relation between the susceptibilities is obtained:

\begin{equation}
	\chi_{T} = - \frac{H}{T} \, \chi_{H} +  \frac{K}{kT^{2}} 
	\left ( \langle cos^{3} \varphi \rangle - \langle cos^{2} \varphi \rangle
	\langle cos \varphi \rangle \right )
\label{Suscept}
\end{equation}

It is hence possible to identify the contributions due to the field
susceptibility (first term in the right hand side in Eq. 
(\ref{Suscept}), and the correction due to the fluctuations of higher order.  In Fig. 
6, both profiles are superimposed : the contribution ($H.\chi_{H}/T$)
is deduced from the measured AMR (red dashed line) and $ a3/I $ is
measured with our protocol.  The deviation is due to the contribution
of the fluctuations of higher order, that tends to reduce the apparent anisotropy (the
minima are pushed toward zero).  This observation will be useful in
order to understand the effect of high current injection in the next
section.

\begin{figure}
\begin{center}
	\includegraphics[height=8cm]{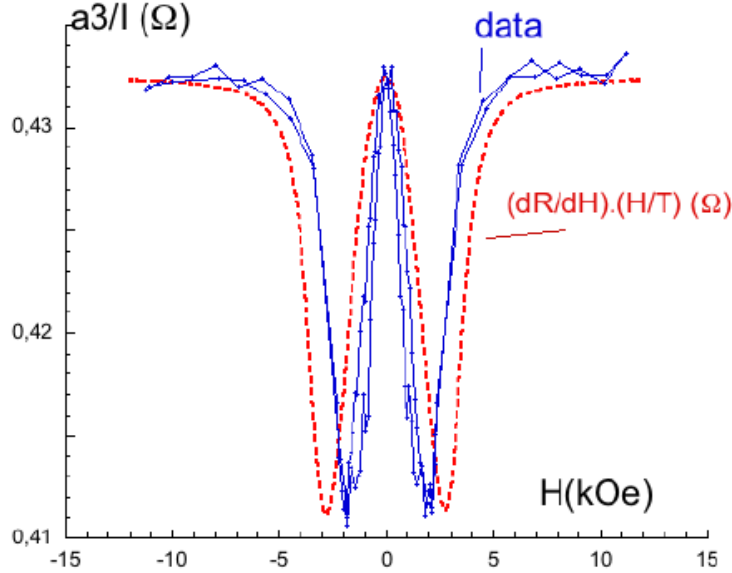}
	\caption{Comparison between the thermal susceptibility (data) and
	the field susceptibility times $H/T$ (dashed line) deduced.  The
	deviation is due to the contribution of the moments of higher
	order (see text)}.
	\label{fig:suscept}
\end{center}
\end{figure}

\subsection{Ferromagnetic Entropy and spin-transfer}

It has been shown that the spin-transfer effect occurs in these samples
while injecting a DC current (or pulsed current) with a density of the
order of $10^{7} A/cm^{2}$ \cite{EPL,PRBKelly,Dubey}.  The
spin-transfer is evidenced by the shift of the switching field, i.e.
the shift of the position of the irreversible jump, proportional to
the current above a given threshold of the order of $10^{7}$
$A/cm^{2}$.  A decrease of the switching field by a factor up to 50 \%
\cite{Dubey} has been observed.  The after-effect measurements (or
ferromagnetic relaxation) show that the effect of the current
injection induces an important modification of the parameters entering
in the N\'eel-Brown activation law (related to the switching field). 
The effective potential barrier - or in stochastic terms the inverse
of the effective temperature - decreases linerarly as a function of
the amplitude of the current.  The effect is strong: 
60 \% variation is obtained between 1.5 and 4 $10^{7}$ $A/cm^{2}$ (see fig.  4,
first paper in Ref \cite{Guittienne}).  Meanwhile, the reversible
part of the hysteresis loop is not significantly modified (see Fig. 
7).  Note that all these features are also observed in spin-valve
structures \cite{PRB03,MSU,Fabian}.  These observations motivated the
interpretation of a fully stochastic process generated by the kinetic
coupling between the spin of conduction electron and the ferromagnetic
order parameter (through the introduction of a relevant Onsager
transport coefficient) \cite{PRB08}.

In this context, the measurements of entropy at equilibrium
(quasi-static states) allows the contribution of ferromagnetic
fluctuations to be quantified and compared to the deterministic
contribution in terms of rotation of the magnetization, as performed 
below.  

\begin{figure}
\begin{center}
	\includegraphics[height=8cm]{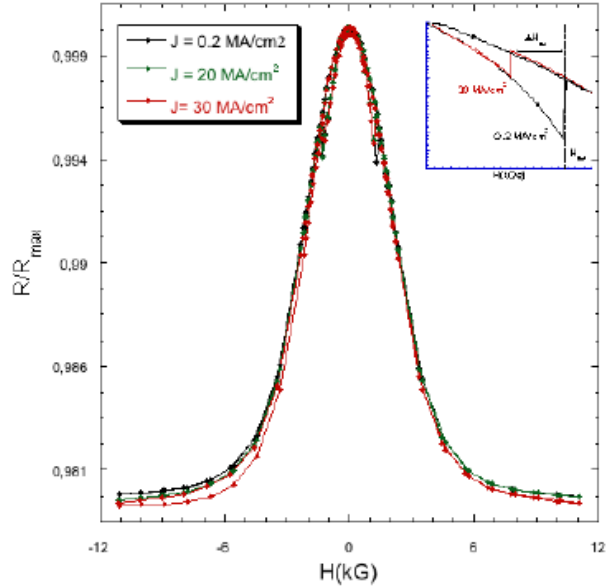}
	\caption{Comparison between AMR Hysteresis loops of reference, and
	under high current injection.  The reversible part of the hysteresis is
	not significantly affected by spin-transfer. Inset: zoom of the 
	irreversible part with the two switching fields}
	\label{fig:Hyst}
\end{center}
\end{figure}

Fig.  7 shows three hysteresis loops normalized to $R_{max}$, with
currents density 0.2 $MA/cm^{2}$, 20 $MA/cm^{2}$ and 40 $MA/cm^{2}$):
the three profiles seem to be superimposed except for the position of
the switching field (not shown here: see
\cite{EPL,Guittienne,PRBKelly,Dubey} for the study of the irreversible
jumps).  The signal corresponding to the rotation of the 
magnetization, if any, is too small to be extracted from the noise in Fig.  7.

On the other hand, Fig.  8 shows that the thermal susceptibility
$a3/I(H)$ as a function of the magnetic field is significantly
modified by injecting high currents.  Since the contribution due to
the Joule effect does not depends on the magnetic field, the variation
observed is due to the magnetic susceptibility.  Fig.  8(a) shows that the
field dependence of the thermal susceptibility is significantly
modified.  Fig.  7 (b) compares the thermal susceptibility of the
magnetoresistance for a weak current (reference current 0.2
$MA/cm^{2}$) and a strong current (45 $MA/cm^{2}$). The difference is due
to the current injection.  In the inset, the
comparison is performed with the two corresponding ferromagnetic
susceptibility deduced from Eq.  (\ref{dRdT1}). 

\begin{figure}
\begin{center}
	\includegraphics[height=10cm]{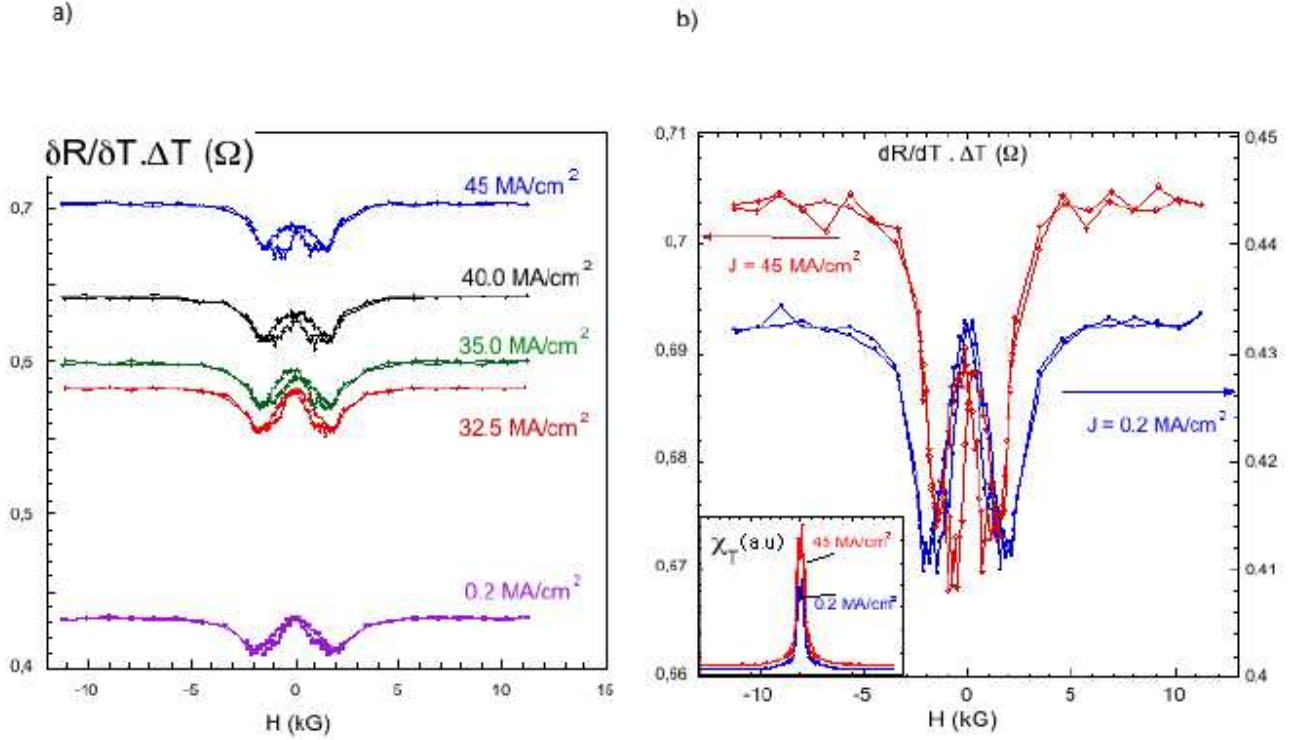}
	\caption{(a) Thermal susceptibility of the magnetoresistance
	(coefficient $a3/I$) as a function of the magnetic field for different 
	current densities.  (b)
	Comparison of the measurements with reference currents $J_{0}=0.2$
	$MA/cm^{2}$ and at high current
 J = 45 $MA/cm^{2}$. Inset: corresponding magnetic
	susceptibilities (deduced from Eq.  (\ref{dRdT1}) and the measured
	AMR), some scale for the magnetic field}
	\label{fig:Chi(I)}
\end{center}
\end{figure}

 According to the following Maxwell equation (see e.g. the discussion 
 developed in the context of magnetocaloric effect 
 \cite{Magnetocaloric,Basso}),
 
\begin{equation}
\delta S = \int_{-H_{sat}}^{H_{sat}} \left (
\frac{\partial M}{\partial T} \right)_{H} dH
\label{Maxwell}
\end{equation}

and recalling that the ferromagnetic system is at equilibrium (only
the reversible branches of the hysteresis are investigated), the
entropy variation due to the high current injection can be deduced
from the measurements of the thermal susceptibility, presented in Fig. 
8. 
We will take as a reference for the entropy of the ferromagnetic
layer the integral of the susceptibility measured at 0.2
$MA/cm^{2}$ in the field range $\pm 1$ Tesla: $S_{ref} \approx$ 8 $10^{-5} JK^{-1} 
kg^{-1}$ (using for Ni
magnetization $M_{s} = 580.  10^{-3}$ $J/Tcm^{3}$ and for Ni density 8.9 
g/$cm^{3}$).

In order to study the entropy generated by spin-transfer, 
i.e. for high current density,
the entropy production is taken from the reference current density 
$J_{0} = 
0,2$ $MA/cm^{2}$ ($I_{0} = 0,02$ mA):

\begin{equation}
\Delta S_{ST}(J) = \int_{-H_{sat}}^{H_{sat}} \left (
\frac{\partial M}{\partial T}(J)- 
\frac{\partial M}{\partial T}(J_{0}) \right)_{H} dH 
\label{Maxwell2}
\end{equation}

The result $\Delta S/S_{ref}$ is plotted in Fig.  8.  The magnetic
field induced by the current has been taken into account in the error
bars as follow: the induced field (about $\pm 5$ $10^{-3}$ $T/mA$, see
discussion in \cite{PRBKelly}) is taking into account with a homogeneous field of
$\pm 0.1$ Tesla added to the integration interval.  The variation is
linear with the current and reaches 60 \% at 4.5 $10^{7}$ $A/cm^{2}$. 
A current threshold is clearly visible and is of the order of 2
$10^{7}$ $A/cm^{2}$.  All those characteristics are quantitatively and
qualitatively that measured for spin-transfer
experiments in Ni nanowires based on activation processes on the
irreversible jumps.

Furthermore, the analysis of the variations due to current injection
(Fig.  8 (b)) shows that the spin-transfer reduces the apparent
anisotropy (i.e. the position of the minima peaks).  The main
contribution is hence apparently due to the fluctuations related to
the moments of high orders (of the type $\langle cos^{3}(\varphi)
\rangle $ or $ \langle cos^{2}(\varphi) \rangle \langle cos(\varphi)
\rangle $. This problem, namely the identification of the specific form of the 
fluctuations, will be investigated in more detail elsewhere.

\begin{figure}
\begin{center}
	\includegraphics[height=10cm]{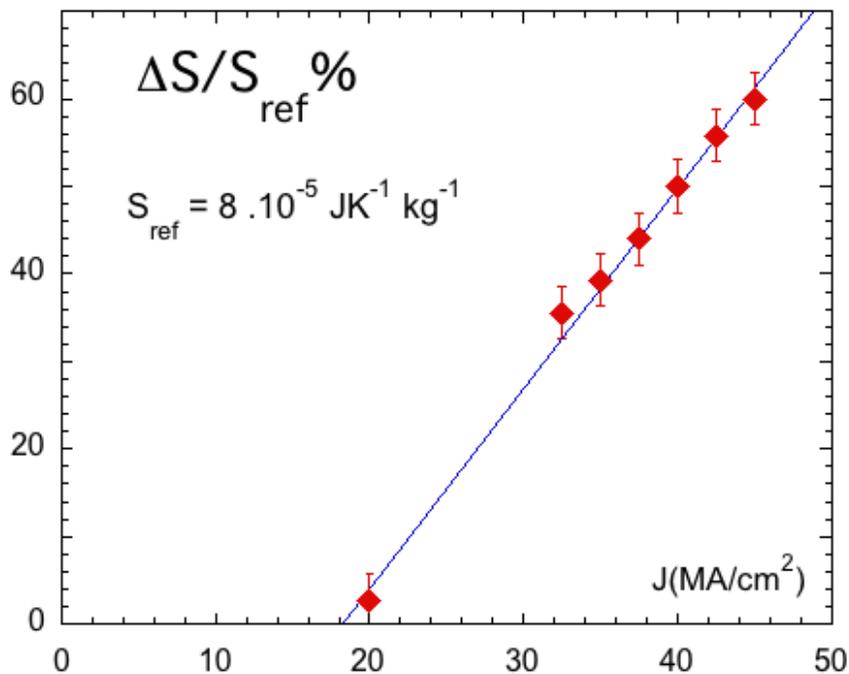}
	\caption{Ferromagnetic entropy production due to current injection as a 
	function of the current density. The entropy is calculated in the 
	field range $\pm H_{sat} = \pm 1T$. The error bars account for the  
	field induced by the current and Joule heating.}
	\label{fig:SI}
\end{center}
\end{figure}

\subsection{Conclusion} 

An experimental protocol has been described that allows the effect of
a high DC current injected in a ferromagnetic nanowire to be
investigated on the reversible branches of the hysteresis loop.  The
protocol is based on the measurements of the thermal magnetic
susceptibility through the magnetoresistance properties and allows
the ferromagnetic fluctuations and the entropy generated by
spin-injections at high current densities to be deduced.

The results show a giant linear increase of the ferromagnetic entropy with
the current injection: up to 60 \% at 4.5 $10^{7}$ $A/cm^{2}$.  A
clear current threshold is also present at about 2 $10^{7}$ $A/cm^2$. 
These characteristics are exactly that previously measured on the
irreversible part of the hysteresis in the case of experimental
protocols based on the activation process (e.g. effective energy
barrier, measure of critical current at given magnetic field, or
measure of the shift of the switching field as a function of the
current).  Consequently, we attribute the observed increase of
entropy measured as a function of the current (Fig. 9) to spin-transfer effect.

Furthermore, the effect of current injection on the equilibrium
(quasi-static) states of the magnetization shows that the
spin-transfer generates strong fluctuations of the magnetization (or
incoherent excitations), instead of rotation of the
magnetization or coherent 
spin-waves (as does the external magnetic torque).  Unfortunately, at this stage
of the study, it is not possible to definitively answer to the
question of whether the spin-transfer is fully stochastic (diffusion
term in the dynamic equation) or if it can be described by a
deterministic term in the dynamical equation.  However, in the last
case, the study shows that the effect of the deterministic torque
generates fluctuations (incoherent excitations), and not coherent
rotation or coherent spin-waves.  

On the other hand, it has been also observed that the most
important contribution of the
fluctuations seem to be generated by statistical moments of higher 
order, beyond the second moment. A more detailed analysis is hence
necessary in order to understand the specificity of the excitations 
produced by the spin-transfer, and to be able to answer the above question about 
the stochasticity, in the framework of the thermodynamic approach 
reported here. 

\subsection{Acknowledgement}
We thank L. C. Sampaio, Renato A Silva, and A. P. Guimar\~aoes
for useful discussions and comments.


\begin{references}
	
	
	\bibitem{liste0} M. Tsoi et al. Phys. Rev. Lett. {\bf 80},  4281 
(1998); J-E. Wegrowe et al. Europhysics lett. {\bf 45} 626 (1999);
F. J. Albert et al. Appl. Phys. Lett. {\bf 77} 3809 (2000);
J. Grollier et al. Appl. Phys. Lett. {\bf 78}, 3663 (2001); E. B. 
Myers et al. Phys. Rev. Lett.  {\bf 89}, 196801 (2002); 
J. Z. Sun et al. Appl. Phys. Lett. {\bf 81}, 2202 (2002); 
J.-E. Wegrowe et al.  Appl. Phys. Lett. {\bf 80}, 3775 (2002);
B. Oezyilmaz et al. Phys. Rev. Lett {\bf 91}, 067203 (2003).

	
\bibitem{Berger} L. Berger, Phys. Rev. B {\bf 54}, 9353 (1996).
	
\bibitem{Slon} J. Slonczewski, J. Magn. Magn. Mat,  {\bf 159} L1 (1996)


\bibitem{JPhys} J.-E. Wegrowe, M. C. Ciornei, H.-J. Drouhin, J. Phys.:
Condens. Matter {\bf 19}, 165213 (2007).

\bibitem{Serpico} C. Serpico, I. D. Mayergoyz, G. Bertotti, M. D'Aquino, R. Bonin, Physica
B {\bf 403}, 282 (2008).

\bibitem{Fukushima} Fukushima et al. J. Appl. Phys. {\bf 99}, 08H706 
(2006), Fukushima et al. IEEE Trans-Mag {\bf 41} 2571 (2005).

\bibitem{Weiss} P. Weiss, A. Piccard C. R. Acad. Sci. (Paris) {\bf 166} 
(1918) 352, 
P. Weiss, R. Forrer, C. R. Acad. Sci., {\bf 178} (1924) 1347. 

\bibitem{WW} W. Wernsdorfer et al Phys. Rev. Lett. {\bf 77} (1996), 1873. 

\bibitem{PRLMoi} J. -E. Wegrowe, D. Kelly,  A. Franck, S. Gilbert,
J.-Ph. Ansermet, Phys. Rev. Lett. {\bf 82} (1999), 3681.

\bibitem{PRBYvan} Y. Jaccard, Ph. Guittienne, 
D. Kelly, J.-E. Wegrowe, J-Ph. Ansermet,, Phy. Rev. B,  {\bf 62} (2000), 1141.

\bibitem{JMMM200} A. Fert, L. Piraux J. Magn. Magn. Mat. {\bf 200} 338 (1999).

\bibitem{EPL} J-E. Wegrowe, D. Kelly, Y. Jaccard, Ph. Guittienne, and 
J.-Ph. Ansermet Europhysics lett. {\bf 45} 626 
(1999), J.-E. Wegrowe, D. Kelly, T. Truong, Ph. Guittienne, J.-Ph. 
Ansermet, Europhysics lett. {\bf 56}, 748 (2001).

\bibitem{Guittienne} Ph.  Guittienne, J.-E. Wegrowe, D. Kelly, and
J.-Ph. Ansermet, IEEE Trans.  Mag.  Magn-{\bf 37}, 2126 (2001),
Ph. Guittienne, L. Gravier, J.-E. Wegrowe, and J.-Ph Ansermet J.
Appl. Phys. {\bf 92}, 2743 (2002), J.-E. Wegrowe, X. Hoffer,
Ph.  Guittienne, A. Fabian, L. Gravier, T. Wade, J-Ph.  Ansermet,
J. Appl. Phys {\bf 91} , 6806 (2002).

\bibitem{PRBKelly} D. Kelly, J. -E. Wegrowe,
Trong-kha Truong, X. Hoffer, Ph.  Guittienne, and J. -Ph. Ansermet.
Phys. Rev.  B {\bf 68}  134425 (2003).

\bibitem{Dubey} J. -E. Wegrowe, M. Dubey, T. Wade,
H. -J. Drouhin, and M. Konczykowski, J. Appl. Phys. {\bf 96} 4490
(2004).

\bibitem{Gravier} L. Gravier, J. -E. Wegrowe, T. Wade, A. Fabian, and
J. -Ph.  Ansermet, IEEE Trans.  Mag.  Mat.  {\bf 38}, 2700 (2002), L.
Gravier, A. Fabian, A. Rudolf, A. Cachin, J. -E. Wegrowe, and J. -Ph. 
Ansermet, J. Magn.  Magn.  Mater.  {\bf 271},153 (2004), L.
Gravier, S. Serrano-Guisan, and J. -Ph.  Ansermet, J. Appl.  Phys. 
{\bf 97}, 10C501 (2005), L. Gravier, S. Sarrano-Guisan, F. Reuse, and 
J.-Ph. Ansermet, Phys. Rev. B, {\bf 73}, 052410 (2006).

\bibitem{Santi} S. Serrano-Guisan, L. Gravier, M. Abid, and
J. -Ph.  Ansermet, J. Appl. Phys. {\bf 99}, 08T108 (2006).

\bibitem{MTEPW} J.-E. Wegrowe, Q. Anh Nguyen, M. Al-Barki, J.-F. Dayen,
T. L. Wade, and H.-J. Drouhin, Phys. Rev. B {\bf 73} 134422, (2006).

\bibitem{Shapira} E. Shapira, A. Tsukernik, and Y. Selzer, 
Nanotechnology {\bf 18}, 485703 (2007).

\bibitem{Myers} E. B. Myers et al. Science {\bf 285}, 867 (1999)

\bibitem{ChenPRL} T. Y. Chen, Y. Ji, M. D. Stiles
and C. L. Chien,  Phys. Rev. Lett. {\bf 93}, 026601 (2004).

\bibitem{PRB08} J.E. Wegrowe, S. M. Santos, M.-C. Ciornei, H.-J.
Drouhin, M. Rubi., Phys.  Rev.  B {\bf 77}, 174408 (2008)

\bibitem{Basso} V. Basso, G. Bertotti, M. LoBue, C.P. Sasso,
J. Magn. Magn. Mat. {\bf 290} (2005) 651.

\bibitem{IEEEMoi} J. -E. Wegrowe, S. Gilbert, D. Kelly, 
 B. Doudin J.-Ph. Ansermet  IEEE Trans. Magn.  {\bf MAG.- 34} (1998), 903.

\bibitem{Travis} T. Wade, J.-E. Wegrowe, Eur. Phys. J. Appl. Phys.
{\bf 29}, 3-22 (2005).

\bibitem{Potter} T. R. McGuire and R. I. Potter, IEEE Trans.  vol {\bf
Mag-11}, 1018 (1975).


\bibitem{Brown} W. F. Brown Jr., {\it Micromagnetics}, Interscience
publishers, 1963.

\bibitem{Felix} S. Serrano-Guisan, G. DiDomenicantonio, M. Abid, 
J.-P. Abid, M. Hillenkamp, L. Gravier, J.-Ph. Ansermet, and Ch. Felix, Nature Mater {\bf 5}, 730 
(2006),

\bibitem{Fullerton} E. E. Fullerton and S. Mangin, Nature Mat. {\bf 7} 
(2008).

\bibitem{PRB03} J.-E. Wegrowe, Phys. Rev. B {\bf 68}, 214414 (2003).

\bibitem{MSU} S. Urazhdin, O. Norman, W. Birge, W. P. Pratt, and J.
Bass, Phys. Rev. Lett. {\bf 92}, 146803 (2003); S. Urazhdin, H.
Kurt, W. P. Pratt, and J. Bass, Appl. Phys. Lett. {\bf 83}, 114
(2003).

\bibitem{Fabian} A. Fabian, C. Terrier, S. Serrano
Guisan, X. Hoffer, M. Dubey, L. Gravier, J.-Ph.  Ansermet, and J.-E.
Wegrowe, Phys.  Rev.  Lett.  {\bf 91}, 257209 (2003).


\bibitem{Magnetocaloric} V. K. Pecharsky and K.A. Gschneidner, Phys. 
Rev. Lett. {\bf 78} 4494 (1997), A. guigui\'ere et al. Phys. Rev. Lett {\bf 
83}, 2262 (1999), K. A. Gschneidner et al. Phys. Rev. Lett. {\bf 85}, 
4190  (2000), J. R. Sun et al., Phys. Rev. Lett. {\bf 85}, 4191 (2000),
F. Casanova et al. Phys. Rev. B {\bf 66}, 100401 (2002)

\bibitem{Superpara} P. Podar, J. Gass, D. J. Rebar, S. Srinath, H. 
Srikanth, S. A. Morison, E. E. Carpenter, J. Magn. Magn. Mat. {\bf 
307}, 227, S. Hariharan and J. Gass, Rev. Adv. 
Mater. Sci. {\bf 10} (2005) 398

\bibitem{Superpara08} V. Franco, K. R. Pirota, V. M. Prida, A. M. J. C. 
Neto, A. Conde, M. Knobel, B. Hernando, and M. Vazquez, Phys. Rev. B 
{\bf 77}, 104434 (2008).

\end{references}
\end{document}